\documentclass{iaus}
\usepackage{natbib}
\usepackage{journals}
\usepackage[latin1]{inputenc}
\usepackage{graphicx}

\title[] %% give here short title %%
{Spheroid ages, kinematics, and BH relations}

\author[Eric Emsellem]   %% give here short author list %%
{Eric Emsellem$^1$%
\affiliation{$^1$Université de Lyon 1, CRAL, CNRS, UMR5574, Observatoire de Lyon; ENS de Lyon, France,
\break email: emsellem@obs.univ-lyon1.fr \\}}

\pubyear{2006}
\volume{235}  %% insert here IAU Symposium No.
\pagerange{119--126}
\date{?? and in revised form ??}
\jname{Galaxy Evolution across the Hubble Time}
\editors{F.Combes \& J. Palous, eds.}
\begin{document}

\maketitle

\begin{abstract}
I very briefly discuss the ages and kinematics of spheroids as well as the black hole relations, via a few
recent and illustrative studies, which include results on the downsizing, scaling laws, angular momentum 
and central massive objects.
\end{abstract}

\firstsection % if your document starts with a section,
\section{Formation, evolution and downsizing}

The present paradigm for the formation and evolution of spheroids accepts
mergers as a major contributor of the game. The intuitive idea that the most
massive systems systematically form last is now clearly revoked by observations
and numerical simulations (see e.g., \citealt{Treu+05}), and "downsizing" became
a fashionable (revived) term. The question is then whether we can or not trace the
formation history of spheroids via observations and models of their stellar
populations and kinematics (e.g., \citealt{ThomasD+05,Cappellari+06,Rettura+06}). 

\section{Ages of present-day early-type galaxies: going red?}

Galaxy evolution in clusters, and more specifically the merger rate, is expected
to be different than in lower density environment. Observations of a sample of
galaxies in low density environment (\citealt{Collobert+06}) have thus shown overall
younger stellar populations though with a greater age spread
(see also \citealt{Kuntschner+02}). The SDSS view (up to $z \sim 0.1$; 
\citealt{Clemens+06}) then suggests that metallicity and
$\alpha$-enhancement are mostly driven by velocity dispersion (mass),
but that only age is really significantly dependent on the environment
(galaxy density; see also \citealt{Nelan+05,Bernardi+06}).
This was confirmed in the study by \cite{GCBW06} 
of the SDSS colour-magnitude and Mg-$\sigma$ relations, which revealed 
that the later is indeed mainly
a mass sequence, a tighter correlation being obtained when
introducing the dynamical mass. The ratio between the stellar mass and the
dynamical mass decreases with mass, which may be interpreted as an increase of
the dark matter contribution for more massive early-type galaxies.
\cite{2006astro.ph..7648B} also showed that galaxy density $\Sigma$ 
could in fact serve as a parameter seconding mass to explain the fraction 
of galaxies in the red sequence. 
These observational results all require a significant role from
an efficient mass dependent feedback mechanism. 

\cite{Faber06} discussed several relative
paths which would transform galaxies from the blue cloud into 
galaxies of the red sequence, including efficient quenching of star formation, 
plus wet and/or dry merging. In this context, stars in the discs and spheroids
should end up with a different formation history, a result emphasised
for nearby early-type galaxies by \cite{Kuntschner+06} via integral-field spectroscopy.
Semi-analytic models (e.g., \citealt{Springel+05}) now include some specific
treatment for quenching via AGN (e.g., \citealt{Croton+06}) which
allows one to follow the star formation history of a large sample of
galaxies with widely different global properties. This led to and confirmed the 
important realisation that the assembly time does not correspond to the formation 
time (of the stars), with massive ellipticals forming their stars
early but being assembled late (\citealt{DeLucia+06}). 

\section{Scaling relations}

Early-type galaxies follow global relations which may be used, when linked
with e.g., information on their stellar population, to constrain
both their internal structure and formation history. 
One of the most powerful relation is the fundamental plane (FP), recently discussed
by \cite{Cappellari+06} who showed that the tilt of the FP with respect to the
law expected from the Virial Theorem is due mostly to an intrinsic variation 
of the mass-to-light ratio, with a maximum contribution of 6\% from
non-homology. The building of state-of-the-art dynamical models also recently allowed
Cappellari et al. (2007, submitted to MNRAS) to suggest a possible link 
between anisotropy and intrinsic flattening. A new quantification of the specific 
angular momentum of such galaxies finally revealed that fast and slow rotators may have 
different formation origins, with an overall trend for slow rotators to be more
massive (Emsellem et al., 2007, submitted to MNRAS). Such information is key
to reconstruct a self-consistent scenario for the formation 
and assembly of early-type galaxies, including mass-dependent criteria 
for the onset of a feedback process and the characteristics of the merged progenitors.
The fact that all slow rotators contain kinematically decoupled cores (KDCs,
Emsellem et al., 2007) may then be explained by the (often underestimated) importance of  
multiple mergers (Bournaud, private communication), although more detailed 
numerical simulations in a cosmogical context (merger tree) are required to confirm this result.

\section{Black hole relations}

Central massive black holes seem to be ubiquitous in nearby galaxies, and 
their mass correlate with various properties 
such as bulge luminosity (\citealt{Magorrian+98}) or mass (\citealt{MH03}), 
stellar velocity dispersion (\citealt{FM00}; \citealt{Gebhardt+00}), 
or luminosity concentration index (\citealt{GECT01}, \citealt{2006astro.ph..7378G}). 
Still, only a few (mostly early-type) galaxies have well-constrained black hole masses
($M_{bh}$),
and progress is clearly required both in terms of more realistic dynamical
modelling and better data. The combined use of integral-field (IF) and high 
spatial resolution data is one step forward (e.g., \citealt{Shapiro+06}).
The recent advent of adaptive-optics driven spectroscopy (\citealt{Houghton+06}) 
and even IFS (Neumayer, in prep.) will also help in this context, 
although such techniques may be restricted for some time to a few specific cases.

The power-law relation between $M_{bh}$ and $\sigma$ has recently been revisited
by Wyithe (2006a, 2006b) who finally found
no significant evidence for a log-quadratic term in the $M_{bh}$-$\sigma$ relation. 
A similar approach was recently applied by \cite{2006astro.ph..7378G} 
on the $M_{bh}$-Sersic~$n$ relation, for which a second order term 
seems to be required. This may significantly influence our view of the high and low mass
ends of the $M_{bh}$ spectrum. An extension of such a relation has been recently
worked out by \cite{ZGZ06} who advocated the existence of a fundamental 
manifold uniting all spheroids present in dark matter halos. 
\cite{2006astro.ph..6739L} also suggest that the $M_{bh}$-$L$ relation may
be in conflict with the $M_{bh}$-$\sigma$ relation, suggesting the latter to
lead to a significant underestimate of the black hole masses 
at the high end (for e.g., brightest cluster galaxies). An alternative view emerges
from a study by \cite{2006astro.ph..9300B} who revealed a bias in the $M_{bh}$-$L$ relation
and favoured the $M_{bh}$-$\sigma$ as a more fundamental relation.
The evolution of the black hole mass with time was studied by \cite{2006astro.ph..7424M}
who estimated that growth via direct accretion is dominant over growth via mergers
only for low mass black holes.
Then, \cite{Ferrarese+06} and \cite{WH06} revealed that central massive (stellar)
objects (CMOs) follow the same relation than massive black holes,
suggesting a unifying mechanism for both nuclei and central black holes
(see also \citealt{2006astro.ph..7378G}).
This may again reflect a mass-dependent phenomenon. It may therefore 
be interesting to look for a link between the proposed AGN-driven 
quenching scheme, the galaxy merger trees and the formation of central mass concentrations.

\section{Conclusions}\label{sec:concl}

In this short paper, I tried to provide a brief
updated view on spheroids. We now have a better understanding 
of the downsizing issue, which should certainly not be interpreted
as an anti-hierarchical process. Environment and mass are key drivers 
of the age and chemical composition of spheroids. A new key parameter 
for galaxy evolution may in fact be found in the amount of baryonic angular momentum.
More work is required to understand the observed relations with
supermassive black hole masses, particularly at the high mass end,
as well as regarding the link between nuclear black holes and CMOs.
Mentioned issues, such as the existence of the blue and red sequences,
the tilt of the FP, the properties of fast and slow rotators, the black hole
(and CMO) relations, all point towards a picture in which galaxy formation and evolution
has been strongly influenced by some efficient, mass-dependent, feedback mechanism.
I would finally like to emphasise the need for a more physically 
motivated quenching scheme, which should allow us to design convincing and tasty
recipes to be implemented in cosmological simulations.

\begin{acknowledgments}
I would like to warmly thank Ryan Houghton, Laura Ferrarese, Harald Kuntschner, Nadine Neumayer,
and Reynier Peletier, for sharing their expertise (and figures!) during the
preparation of this review, Michele Cappellari for constructive and thoughtful comments, 
and Alister Graham for useful input.
\end{acknowledgments}
%%\bibliographystyle{astron}
%%\bibliography{ref}
%% \begin{thebibliography}{}
%% \bibitem[Ferrarese et al.(2006)]{2006ApJ...644L..21F} Ferrarese, L., et 
%% al.\ 2006, \textit{ApJ}, 644, L21 
%% \end{thebibliography}

\begin{discussion}

\discuss{Mike Dopita}{I am glad you emphasised the need for physical models of AGN feedback. 
The observations of high-$z$ radio-galaxies show that this feedback leads to violent star formation along
with the matter ejection. Do you know of any observations that could reveal
the fossil evidence of this star-formation episode?}

\discuss{Eric Emsellem}{Not really, but I must admit I may not be the right (competent) person to answer that
question. My gut feeling is that, on one hand, we should not expect much fossil evidence for such episodes 
in the local Universe, first because the information has been "smoothed" out since then, but 
also because it would be extremely difficult to disentangle the contribution from AGN-induced star 
formation from other mechanisms. On the other hand, a specific and more direct signature of such violent episodes may exist
(since it does make a difference in the shaping of the hosts in terms of energy budget and colour/stellar population).
To my knowledge, there is so far no such obvious signature to look for.}
\end{discussion}

\end{document}